\documentclass[12pt]{article}
\usepackage{palatino,cite,epsfig}
\usepackage{graphics,graphicx}

\newlength{\capindent}
\newlength{\capwidth}
\newlength{\figwidth}
\newcommand{\icaption}[2][!*!,!]{\hspace*{\capindent}%
  \begin{minipage}{\capwidth}
    \ifthenelse{\equal{#1}{!*!,!}}%
      {\caption{#2}}%
      {\caption[#1]{#2}}
  \end{minipage}}
\def\a34{\cos\alpha_{34}}

\def\AFB{\mathrm{\rm A_{FB}\;}}

\def\xi{x_{i}}

\def\wangle{\mathrm{sin^{2}\theta_{eff}^{lept}(M^2_Z)}}
\newcommand{\GeV}{\ensuremath{\mathrm{Ge\kern -0.12em V}}}
\newcommand{\TeV}{\ensuremath{\mathrm{Te\kern -0.12em V}}}

\newcommand {\Be}{\begin{equation}}
\newcommand {\Ee}{\end{equation}}

\newcommand {\Tabref}[1]{Table~\ref{tab:#1}}

\begin{document}

\vspace{1cm}

\begin{flushright} 
hep-ph/0305126 \\
May 2003
\end{flushright}
\vspace{0.5cm}

\begin{center}

{\Large\bf Study of Parton Density Function Uncertainties with LHAPDF and PYTHIA at LHC}

\vspace{0.4cm}

{\bf Contribution to LHC / LC Study Group Working Document%
\footnote{
For further informations, see {\tt
http://www.ippp.dur.ac.uk/$\sim$georg/lhclc/ }. For questions and comments,
please contact {\tt Georg.Weiglein@durham.ac.uk }.}
}
 
\vspace{1cm}

{\sc 
Dimitri Bourilkov
}

{\small bourilkov@mailaps.org
}

\vspace*{1cm}

{\sl
    University of Florida\\
    Gainesville, FL, USA

\vspace*{0.4cm}
}
\end{center}

\vspace*{1cm}

\begin{abstract}
The experimental errors in current and future hadron colliders are expected 
to decrease to a level that will challenge the uncertainties in the theoretical
calculations. One important component in the prediction uncertainties comes from the
Parton Density Functions of the (anti)proton. In this work we develop an interface
from the Les Houches Accord Parton Density Functions (LHAPDF) package to the very
popular Monte Carlo generator {\tt PYTHIA} version 6.2. Then we proceed to 
estimate the PDF uncertainties for the production of Drell-Yan pairs 
from the Z pole to masses above 1~$\TeV$ and for Higgs bosons at LHC.
The measurement of the electro-weak mixing angle at LHC as a particularly
difficult case is studied.
\end{abstract}

{\it Work presented at the LHC/LC Study Group (CERN 5 July 2002, 14 February and 9 May 2003)}



\section{LHAPDF Interface to {\tt PYTHIA} 6.2}

A hadron collider can be viewed as a wide band parton beam machine.
To make predictions, the parton cross sections are folded with the
parton density functions (PDF):
\begin{equation}
\rm \frac{d\ \sigma}{d\; variable}[pp\rightarrow X] \sim \sum_{ij} 
\, \left(f_{i/p}(x_1)  f_{j/p}(x_2) +(i \leftrightarrow j)\right)\,  \hat{\sigma}  
\label{llpdf1}
\end{equation}
$\rm \hat{\sigma}$ - cross section for the partonic subprocess $ij\rightarrow X$\\
$\rm x_1$, $\rm x_2$ - parton momentum fractions,\\
$\rm f_{i/p(\bar{p})}(x_i)$ - probability to find a parton $i$ with momentum
fraction $x_i$ in the (anti)proton.

A long standing problem when performing such calculations is what is the
uncertainty of the results coming from our limited knowledge of the PDFs,
even if the parton cross section $\rm \hat{\sigma}$ is known very
precisely. The constantly increasing computing power allows to address
this problem in new ways.

The Les Houches Accord Parton Density Function interface was conceived at the 
Les Houches 2001 workshop in the PDF working group to enable the usage of
Parton Density Functions with uncertainties in a uniform manner~\cite{lhapdf}.
The LHAPDF code is designed to work with PDF sets. 
In this approach a ``fit''
to the data is no longer described by a single PDF, but by a PDF set 
consisting of many individual PDF members. Calculating an observable for all
the PDF members (by repeating the Monte Carlo generation N times, consuming
lots of CPU time), one can estimate the uncertainty on the observable coming
from the limits in our PDF knowledge. The LHAPDF code was built with this
in mind and manipulates PDF sets.

Some sets available in LHAPDF are:
\begin{itemize}
\item CTEQ6~\cite{Pumplin:2002vw} with 40 members: uses $2\cdot N_{PDF}$
      eigenvectors for
      twenty PDF parameters, selected to represent the PDF variation;
      to estimate an uncertainty corresponding to $\sim$ 90 \% CL
      the following sum is constructed:
$$\Delta X = \frac{1}{2}\sqrt{\sum_{i = 1}^{N_{PDF}} [X_{i}(S_+) - X_{i}(S_-)]^2}$$
\item MRST2001~\cite{Martin:2001es} with 2 members
\item Fermi2002~\cite{Giele:2001mr} with 100(1000) members: standard statistical
      procedures are
      used to estimate the standard deviation from the set
\item two $Q^2$ evolution codes: EVLCTEQ v1 for CTEQ6 and QCDNUM v 16.12 for
      all other sets.
\end{itemize}

The interface developed by the author~\footnote{Available on request from
bourilkov@mailaps.org.} enables the use of modern PDFs,
as provided by the Les Houches Accord Parton Density Functions (LHAPDF) 
package, from the very popular Monte Carlo generator 
{\tt PYTHIA}~\cite{pitia} (version 6.2).
The user can select for a given generation the
PDF set and the member in the set with standard {\tt PYTHIA} data cards. One
member of a PDF set is fixed at initialization time and used throughout one
run. The user can repeat the generation as many times as needed (or feasible!)
in order to compute not only the central values (from the member representing
the ``best'' fit in a set), but also the uncertainties of various observables
(e.g. total or differential cross sections, forward-backward asymmetries etc.).

\subsection{Consistency Checks}

To check our implementation, we have computed some cross sections
using different PDFs. The {\tt PYTHIA} default is CTEQ5L. The same
PDF is implemented in PDFLIB~\cite{pdflib} from the CERN library.
The cross sections for Drell-Yan pairs ($e^+e^-$) with {\tt PYTHIA} 6.206,
generating each time samples of 100000 events, are shown in~\Tabref{consistency}.

\begin{table}[htb]
{\large
\caption{Cross sections for Drell-Yan pairs ($e^+e^-$) with {\tt PYTHIA} 6.206.
The errors shown are the statistical errors of the Monte-Carlo generation.}
\label{tab:consistency}
  \begin{center}
\begin{tabular}{|l|l|c|}
\hline
PDF set   &  Comment                    &  xsec         \\
\hline
             \multicolumn{3}{|c|}{$81 < M < 101$ GeV}    \\
\hline
CTEQ5L    &  PYTHIA internal            &  1516 $\pm$ 5 pb      \\
CTEQ5L    &  PDFLIB                     &  1536 $\pm$ 5 pb      \\
CTEQ6     &  LHAPDF                     &  1564 $\pm$ 5 pb      \\
MRST2001  &  LHAPDF                     &  1591 $\pm$ 5 pb      \\
Fermi2002 &  LHAPDF                     &  1299 $\pm$ 4 pb      \\
\hline
             \multicolumn{3}{|c|}{$M > 1000$ GeV}    \\
\hline
CTEQ5L    &  PYTHIA internal            &  6.58 $\pm$ 0.02 fb      \\
CTEQ5L    &  PDFLIB                     &  6.68 $\pm$ 0.02 fb      \\
CTEQ6     &  LHAPDF                     &  6.76 $\pm$ 0.02 fb      \\
MRST2001  &  LHAPDF                     &  7.09 $\pm$ 0.02 fb      \\
Fermi2002 &  LHAPDF                     &  7.94 $\pm$ 0.03 fb      \\
\hline
    \end{tabular}
  \end{center}
}
\end{table}

We have performed the calculations at the Z pole and for very high masses
above 1~$\TeV$ in order to check the PDFs in different $x$ regions.
The probed values are summarized in~\Tabref{xmasrap}.
Good agreement between the CTEQ6 and MRST2001 sets
for a large $Q^2$ range is observed, taking into account that two
different evolution codes are used. The results are close to the
cross sections obtained with CTEQ5L. Even for the same PDF we
observe a minor difference between the internal {\tt PYTHIA}
implementation and the PDFLIB one.
Each PDF set has a corresponding ``best fit'' or preferred value
for the strong coupling constant $\alpha_s$ and for the number of flavors.
Our interface has not been optimized yet to give the best values for
each set, and this will happen in the next release after detailed
investigation. The necessary code is already foreseen in the current
version and the values can be set by the user.

\begin{table}[htb]
  \begin{center}
\caption{x$_1$ and x$_2$ for different masses and rapidities.}
\label{tab:xmasrap}
\vskip0.2cm
\begin{tabular}{|c|c|c|c|}
\hline
   y      &     0         &     2       &      4        \\
\hline
             \multicolumn{4}{|c|}{M = 91.2 GeV}          \\
\hline
$\rm x_1$ &   0.0065      &   0.0481    &    0.3557     \\
$\rm x_2$ &   0.0065      &   0.0009    &    0.0001     \\
\hline
             \multicolumn{4}{|c|}{M = 200  GeV}          \\
\hline
$\rm x_1$ &   0.0143      &   0.1056    &    0.7800     \\
$\rm x_2$ &   0.0143      &   0.0019    &    0.0003     \\
\hline
             \multicolumn{4}{|c|}{M = 1000 GeV}          \\
\hline
$\rm x_1$ &   0.0714      &   0.5278    &       -       \\
$\rm x_2$ &   0.0714      &   0.0097    &       -       \\
\hline
    \end{tabular}
  \end{center}
\end{table}

\section{PDF Uncertainties}

As discussed above, the uncertainty on an observable is estimated by
repeating the calculation many times using in turn all the members of
a PDF set. As we are generating events with {\tt PYTHIA} here,
we have to be careful not to overestimate the PDF uncertainty.
Namely, the observable under study will have a natural variation
coming from the limited Monte-Carlo statistic. This complication
is taken into account by:
\begin{itemize}
 \item when possible generating large enough samples so that the MC error is
       an order of magnitude lower than the PDF contributions
 \item when this is not feasible, unfolding the MC contribution from
       the total variation to extract the ``pure'' PDF part; if the
       observed variation is equal to the expected from the Monte-Carlo
       statistic, this means that we cannot estimate the PDF uncertainty
       and can only give an upper limit - improving in such a situation
       entails generating huge samples.
\end{itemize}

\subsection{Results at the Z Pole}

We start with cross sections for Drell-Yan pairs ($e^+e^-$) with {\tt PYTHIA} 6.206.
This time a cut on the rapidity~$<\ 2.5$ is applied, corresponding to the barrel and
endcaps of a LHC detector. Samples of 1200000 events are generated for each member of
a set. This process probes the (anti)quark PDFs.

\begin{table}[htb]
{\large
\caption{Cross sections for Drell-Yan pairs ($e^+e^-$) with {\tt PYTHIA} 6.206.
The errors shown are the PDF uncertainties.}
\label{tab:zpole}
  \begin{center}
\begin{tabular}{|l|l|r|c|}
\hline
PDF set   &  Comment                    &  xsec [pb] & PDF uncertainty \%    \\
\hline
             \multicolumn{4}{|c|}{$81 < M < 101$ GeV}    \\
\hline
CTEQ6     &  LHAPDF                     &  1065 $\pm$ 46  &  4.4 \\
MRST2001  &  LHAPDF                     &  1091 $\pm$ ... &      \\
Fermi2002 &  LHAPDF                     &   853 $\pm$ 18  &  2.2 \\
\hline
    \end{tabular}
  \end{center}
}
\end{table}

Again a good agreement between the CTEQ6 and MRST2001 sets is observed.
If we take into account that in the CTEQ6 approach the estimate
corresponds to $\sim$ 90 \% CL, then at the one standard deviation
level the PDF uncertainties are small ($\sim$ 2--3 \%) and the two methods
of estimation give close values. While the CTEQ6 and Fermi2002 sets agree
on the uncertainties, the difference in their central values requires
further study.

\subsection{Higgs Production}

The cross section for $gg \rightarrow H$ probes the gluon PDFs, so this
channel is complementary to the case considered above.
The CTEQ6 set is used. Samples of 100000 events are generated for each
member of the set and for each value of the Higgs mass, as shown in~~\Tabref{higgs}.

\begin{table}[htb]
{\large
\caption{Cross sections for Higgs production in $gg \rightarrow H$ with {\tt PYTHIA} 6.206.
The errors shown are the PDF uncertainties.}
\label{tab:higgs}
  \begin{center}
\begin{tabular}{|c|r|c|}
\hline
Higgs mass [GeV]  &  xsec [pb] & PDF uncertainty \%    \\
\hline
 120      & 16.26 $\pm$ 0.62&  3.8 \\
 160      & 10.46 $\pm$ 0.32&  3.1 \\
 200      &  7.41 $\pm$ 0.19&  2.6 \\
\hline
    \end{tabular}
  \end{center}
}
\end{table}

We can observe falling PDF uncertainties for higher Higgs masses.
They are always below 4 \%, and we should take them with a grain of salt,
as for this process extrapolations are used for very low $x$, and the PDF
uncertainties are not known so precisely there.

\section{Determination of $\wangle$ at LHC}

As studied for the LHC SM workshop (see~\cite{Haywood:1999qg,dbcmsnote:2000}
and references therein), the huge statistics of lepton pairs from
Z decays can open the possibility to perform a competitive
measurement of the electroweak mixing angle $\wangle$ at LHC,
provided that we can control to the needed precision the 
proton PDFs. This question was left open, and the tools developed
here allow to address it again.

For this end a very precise measurement of the forward-backward asymmetry
at the Z pole is needed. 
For a ($\rm x_1 \geq x_2$) pair of partons producing the Z
we have 4 combinations of {\em up-} or {\em down-}type
quarks initiating the interaction:
$u\bar u, \bar u u, d\bar d, \bar d d$.
In $pp$ collisions the antiquarks come always from the sea and the quarks
can have valence or sea origin.
Going to higher rapidities increases the difference between
x$_1$ and x$_2$ and hence the probability that the first quark is
a valence one. This allows a measurement of the forward-backward asymmetry
even for the symmetric initial $pp$ state.
At the TEVATRON the proton-antiproton configuration provides a
natural label for the quark or anti-quark.

From this discussion it is clear that going to highest rapidities
is helpful at LHC. 
A central detector with $|\eta| < 2.5$ has a much reduced sensitivity
compared to the TEVATRON.
A competitive measurement using $\sim$~100 fb$^{-1}$ at LHC starts at
\mbox{$\Delta\wangle = 0.0005$} per channel, or
\mbox{$\Delta\wangle = 0.00025$} for 2 x 2 channels (electrons and muons)
and experiments. The best single measurement today from SLD~\cite{Abe:2000dq}
using the left-right asymmetry gives \mbox{$\Delta\wangle = 0.00026$}.

In the following we outline a possible strategy to do the measurement.
Events are kept for analysis by applying the following cuts for both leptons:
\begin{itemize}
 \item pseudorapidity $\rm |\eta| < 2.5$
 \item transverse momentum $\rm p_T > 10$ GeV
\end{itemize}
which cover the barrel and endcaps of a typical LHC detector. The backgrounds
for these final states are low and can be suppressed further by isolation cuts.
The dependence of the asymmetry on the measured angle is given by:
$$\frac{\Delta(A_{FB})}{\Delta(\wangle)} = k.$$
Not knowing the quark direction at LHC leads to values $k < 1$
i.e. reduced sensitivity.
Samples of $\sim$~110 million events (after cuts) for different values of $\wangle$
are generated and the results shown in~\Tabref{zafb}.

\begin{table}[htb]
{\large
\caption{Asymmetry measurement at the Z pole at LHC. The coefficient $k$,
the required error on $\AFB$ for a measurement with precision
\mbox{$\Delta\wangle = 0.0005$} in one channel and one experiment,
the upper limit on the PDF uncertainty
and the number of events needed are shown for three rapidity
intervals. In the last row the expected events in one channel for
one experiment with luminosity 100 fb$^{-1}$ and detection efficiency
100 \% in the acceptance region are given.}
\label{tab:zafb}
  \begin{center}
\begin{tabular}{|l|r|r|r|}
\hline
Rapidity            &  0.0 - 0.8 &  0.8 - 1.6   &  1.6 - 2.4  \\
\hline
k                   &   -0.021   &    -0.38     &     -0.59   \\
\hline
$\Delta(\AFB)$      & 0.00001    &   0.00019    &   0.000295  \\
\hline
PDF uncertainty     &$<$ 0.00048 &$<$ 0.00053   &$<$ 0.000820 \\
\hline
Events needed       &  10000 M   &   27.7 M     &   11.5 M    \\
\hline
Events in one       &            &              &             \\
ch/exp 100 fb$^{-1}$&   30.7 M   &   25.0 M     &   10.5 M    \\
\hline
    \end{tabular}
  \end{center}
}
\end{table}

We may perform two independent measurements in the rapidity intervals
0.8--1.6 and 1.6--2.4 in order to control the experimental systematics.
From the table we see that 27.7(11.5) x 10$^6$ events in the two
intervals are needed, which are available from one channel
in one experiment for $\sim$~110 fb$^{-1}$ if the detection efficiency in
the acceptance region is 100 \% (for lower efficiency the luminosity has
to be rescaled accordingly). Clearly the statistical error is not a problem,
and we have two independent measurements for each channel.
From a big run with CTEQ6 - 40 x 15 x 10$^6$ events,
we have determined the upper limits at 90 \% CL of the PDF uncertainties
for different rapidity intervals, as shown in~\Tabref{zafb}.
Even with this statistics we have not reached a systematic show-stopper,
so the question remains open:
it is possible that the PDF uncertainty will not be the limiting
factor, but runs with huge samples are needed to determine it reliably.

\appendix
\section{Appendix}

    The interface is designed along the lines of the existing interface
from PYTHIA to PDFLIB (see e.g. CERNLIB).

\begin{enumerate}
    \item If the user is not interested in LHAPDF, standard PYTHIA can be used
OR the file
    
       pythia6206.f

from the repository. This file has been modified as follows: two PYTHIA
routines (PYINIT, PYPDFU) will perform additional actions if MSTP(52)=3.
And two dummy
routines LHALAUNCH and LHAPARD are added close to the end of the file.
So if the user does not set MSTP(52)=3, nothing changes. And if the
user just happens to set MSTP(52)=3, the dummy routines will be called,
they will issue an error message and stop the execution. For using LHAPDF
please see 2).

    \item If the user is interested in LHAPDF, the file
    
       pythia6206lha.f

from the repository should be used. Compared to pythia6206.f the only change
is that the two dummy routines LHALAUNCH and LHAPARD have been taken out.
Now the user has to link in addition the library

       libLHAPDF.a

which is built using the file Makefile (just give the command make at
the UNIX prompt).
\end{enumerate}

       Some  files from LHAPDF have been modified for 
technical reasons and to facilitate ease of use and transparency for
the user. In the original LHAPDF code there are two evolution codes:
EVLCTEQ used with the CTEQ PDFs, and QCDNUM for other PDFs. The user
is expected to build the library corresponding to ONE of them and
link it. So the choice of PDF cannot be done at runtime and to use
the second option one has to rebuild the library and relink.

       By modifying the code, I have simplified the use: the library
is built only ONCE, and contains the two evolution codes. To use LHAPDF
one sets MSTP(52)=3. The user can select the PDF set and the member of 
the set at runtime with MSTP(51):
\begin{itemize}
\item          CTEQ6 (\verb{file cteq6.LHpdf{):\\
    MSTP(51) = 1000--1040 (1000 - best fit, 1001-1040 - PDF members)
\item          MRST2001 (\verb{file MRST2001.LHpdf{):\\
    MSTP(51) = 2000--2002 (2000 - best fit, 2001-2002 - PDF members)
\item          MRST98   (\verb{file MRST98.LHpdf{):\\
    MSTP(51) = 2100--2103 (2100 - best fit, 2101-2103 - PDF members)
\item          Fermi2002  (\verb{file Fermi2002_100.LHpdf{):\\
    MSTP(51) = 3000--3100 (3000 - best fit, 3001-3100 - PDF members)
\item          Fermi2002  (\verb{file Fermi2002_1000.LHpdf{):\\
    MSTP(51) = 4000--5000 (4000 - best fit, 4001-5000 - PDF members).
\end{itemize}

The PDF sets are stored for convenience in the subdir PDFsets (can be
downloaded from the LHAPDF page). The routine LHALAUNCH is hardcoded
to fetch them from this subdir - if the user decides to store them
elsewhere, the code should be modified to reflect the new location.

       Each PDF set has a corresponding ``best fit'' or preferred value
for the strong coupling constant $\alpha_s$ and for the number of flavors.
The interface has not been optimized yet to give the best values for
each set, and this will happen in the next release after detailed
investigation. The necessary code is already foreseen in the current
version and the values are set in LHALAUNCH.

       ENJOY!


\subsection*{Acknowledgments}

The author would like to thank Albert De Roeck for encouraging this study,
Walter Giele and Joey Huston for useful discussions.
The large scale calculations are performed on the HEE computer farm at the
University of Florida using the CONDOR scheduler~\cite{condor}, a component of
the emerging grid middleware. In total 1200 million {\tt PYTHIA} events
are generated.
This work is supported by the United States National Science Foundation
under grants NSF ITR-0086044 and NSF PHY-0122557.



\end{document}